# Ultrafast and energy-efficient all-optical switching with graphene-loaded deep-subwavelength plasmonic waveguides

Masaaki Ono[1,2]*, Masanori Hata[2,3], Masato Tsunekawa[2,3], Kengo Nozaki[1,2], Hisashi Sumikura[1,2], Hisashi Chiba[2,3], and Masaya Notomi[1,2,3]*

[1]Nanophotonics Center, NTT Corporation, 3-1, Morinosato Wakamiya, Atsugi, Kanagawa, 243-0198, Japan

[2]NTT Basic Research Laboratories, NTT Corporation, 3-1, Morinosato Wakamiya, Atsugi, Kanagawa, 243-0198, Japan

[3]Department of Physics, Tokyo Institute of Technology, 2-12-1, Ookayama, Meguro-ku, Tokyo, 152-8550, Japan

*e-mail: masaaki.ono.bm@hco.ntt.co.jp; notomi@phys.titech.ac.jp

KEYWORDS: all-optical switch, plasmonic waveguides, graphene, photonic integration, plasmonics, silicon photonics

**ABSTRACT**

All-optical switches have attracted attention because they can potentially overcome the speed limitation of electric switches. However, ultrafast, energy-efficient all-optical switches have been challenging to realize due to the intrinsically small optical nonlinearity in existing materials. As a solution, we propose graphene-loaded deep-subwavelength plasmonic waveguides (30 × 20 nm$^2$). Thanks to extreme light confinement, we have significantly enhanced optical nonlinear absorption in graphene, and achieved ultrafast all-optical switching with a switching energy of 35 fJ and a switching time of 260 fs. The switching energy is four orders of magnitudes smaller than that in previous graphene-based devices and is the smallest value ever reported for any all-optical switch operating at a few picoseconds or less. This device can be efficiently connected to conventional Si waveguides and employed in Si photonic integrated circuits. We believe that this graphene-based device will pave the way towards on-chip ultrafast and energy-efficient photonic processing.

All-optical switching has been attracting strong attention for decades because it can potentially overcome the speed limitation imposed by electrical switching devices[1,2]. Indeed, ultrafast switching in less than 1 ps has been demonstrated in a number of systems by using fast optical nonlinearity, such as the Kerr effect[3-10]. However, these all-optical switching devices require a large driving energy of typically pJ to nJ per pulse, which is a barrier to most applications. Although the energy can be reduced by sacrificing speed, there is a general trade-off between speed and energy (see Fig. S1). This problem is primarily attributed to the intrinsically small optical nonlinearity in existing materials[11]. Recently, it was demonstrated that strong light confinement can break this trade-off, and a driving energy of 400-600 aJ was achieved for a switching time of 20-35 ps by using photonic crystal nanocavities with carrier-induced optical nonlinear dispersion[12]. However, it is still difficult to achieve an ultrafast switching regime (less than ps) with this scenario because the switching speed is severely limited by the cavity photon and carrier lifetimes. It is possible to reduce the cavity $Q$ to shorten the photon lifetime, but much a larger nonlinear shift is needed, which is not available with existing materials. Hence, both are more or less fundamental limitations.

In this study, we employ another approach to break this barrier. Here we choose a nanowaveguide with nonlinear absorption (Fig. 1a), instead of a nanocavity with nonlinear dispersion. The first key point of this study is that we employ graphene as a nonlinear saturable absorption element[13,14], in which an opaque medium becomes transparent with optical pulse excitation as a result of Pauli blocking caused by photo-excited carriers[15]. This mechanism does not need a high-$Q$ cavity, and the driving energy is mostly determined by the number of carriers needed to make the waveguide (WG) transparent. Although there have been a vast number of studies related to saturable absorption[16,17], the speed is generally limited because of the long photo-generated carrier relaxation time. In contrast, the relaxation time of photo-generated carriers in graphene is extremely short (10 fs to several ps)[18-21] because of its intriguing gapless band structure and strong phonon scattering. In addition, graphene exhibits very large and flat absorption over a wide wavelength (λ) range, which is also attributed to their unique Dirac-cone bands[22,23]. It was also reported that the saturation power of graphene is one order of magnitude smaller than that of conventional semiconductors and carbon nanotubes[24]. These characteristics are promising for the nonlinear absorption elements of broadband and ultrafast all-optical switches.

However, graphene is too thin to couple efficiently with light. In fact, there have been many graphene-loaded photonic devices[25-32], but they are relatively long (> 40 μm for a Si-waveguide-based structure[25]), which is unsuitable for reducing the driving energy. For example, an all-optical modulator was demonstrated by employing a graphene-wrapped microfibre with an operating speed of 2.2 ps, but the device is 20-μm long and requires 200 pJ at a 1.4-dB contrast[33]. This makes our second ingredient important, namely a deep-sub-λ plasmonic metal-insulator-metal (MIM) WG[34,35]. As described later, we can significantly increase light-matter interactions in graphene by placing it on an ultranarrow MIM-WG (30 × 20 nm$^2$), thus enabling us to shorten the device length to 4 μm and to reduce the effective graphene volume. The strong light confinement indicates that the energy required for saturable absorption should be greatly reduced. As a rough estimation, to fill all the electron states of 20-nm × 4-μm graphene to 0.4 eV and thus induce saturable absorption at 1550 nm, we need $10^4$ photons, corresponding to only 1.2 fJ. The last remaining problem is coupling light into an MIM-WG. This had been a serious issue with MIM devices, but we have solved this problem by employing a very efficient and broadband plasmonic mode converter between MIM- and conventional Si-WGs[36].

Our proposed switch, which combines all the above approaches, is shown in Figs. 1a and 1b, and consists of graphene-loaded MIM-WGs equipped with plasmonic mode converters connected to Si-WGs. We first present our theoretical design for our device concept. Then, we describe the fabrication of our proposed graphene-loaded MIM devices and their experimental characterizations. Finally, we describe our demonstration of the first all-optical switching in the femtosecond and femtojoule regime in any type of device.

**Design**

To make graphene-loaded WG devices energy efficient, it is essential to increase the effective absorbance of graphene by using nanophotonic light confinement. We employed the finite-element method to calculate the optical modes of graphene-loaded devices (see Methods for calculation details). Figure 1c shows the calculated field profile ($|E|^2$) of the eigenmode at $\lambda$ = 1550 nm with an air slot width ($w_{slot}$) of 30 nm and an Au thickness ($t_{Au}$) of 20 nm, which is fundamental even mode in terms of the transverse electric field[34,37]. Note that for an MIM-based device the light field is strongly confined in the air slot whose area is as small as $\lambda^2/4000$. This extreme light confinement around graphene leads to a large increase in absorbance. We estimate the absorption coefficient of graphene $A_G$ from the calculated complex propagation constant with and without graphene, and the estimated $A_G$ is 2.0 dB/μm. As a comparison, we also estimate $A_G$ for a graphene-loaded Si-wire WG with a 400 × 200 nm$^2$ cross section (the field profile is shown in Fig. S2a). The estimated $A_G$ is 0.089 dB/μm, which represents a 22-fold enhancement of $A_G$ in the MIM-WG device compared with a Si-wire WG device.

Figure 1d shows the calculated $A_G$ when varying $w_{slot}$ and $t_{Au}$. This plot shows that $A_G$ is enhanced as the dimension of the air slot ($w_{slot}$ and $t_{Au}$) decreases, which is understood as the field concentration effect (the field profiles are shown in Figs. S2b-e)[38]. It is worth noting that the MIM size should really be in the deep sub-λ regime (< 50 nm) to achieve sufficiently large absorbance enhancement. We also examined the MIM-WG loss $A_{MIM}$, which is inevitable in MIM-WGs, and estimated the ratio $A_G/A_{MIM}$ (Supplementary S4). This ratio increases as the slot size decreases, indicating that deep-sub-λ slots are also advantageous in terms of the MIM-WG loss.

So far, we have considered linear absorption enhancement, but the light concentration affects the nonlinear response even more strongly. Here, we estimate the average field intensity at the graphene position in the MIM-WG slot, and define the field enhancement factor $EF$ against that for the Si-wire WG structure (Supplementary S3). $EF$ increases as the slot size decreases, and a 310-fold enhancement is obtained at $w_{slot}$ = 30 nm and $t_{Au}$ = 20 nm (Fig. 1e). As a result, the saturation power will be greatly reduced.

To enjoy this merit, the coupling efficiency into MIM-WGs is crucial because it is not easy to directly couple light into MIM-WGs from outside. We solve this problem by adopting a plasmonic mode converter[36] as shown in Fig. 1a. Although the MIM-WG and Si-WG have fundamentally different modes in terms of size and characteristics, our mode converter enables efficient and broadband full three-dimensional mode conversion with just a two-dimensional laterally tapered structure. The simulated coupling loss is as small as 1.2 dB. Therefore, we can expect a large overall reduction in the switching energy over a broadband λ range.

**Experiment**

We fabricated WG structures by electron beam lithography, dry etching, and lift-off processes. We transferred graphene onto the MIM-WGs, and patterned the graphene using reactive ion etching (see Methods for fabrication details). Figure 2a shows a scanning electron microscope (SEM) image around the plasmonic mode converter, and the metal part was successfully fabricated despite there being thick Si wire (400 × 200 nm$^2$) in the vicinity of the metal part. Figures 2c and 2d show SEM images of the graphene-loaded MIM-WGs with different graphene lengths $l_G$ (Fig. 2b). We confirmed that the performance of the MIM-WG and mode converters was almost the same as in our previous work[36]. In fact, we achieved almost the same coupling efficiency (−1.7 dB), even though the smallest slot size is reduced to 30 × 20 nm$^2$ from 50 × 20 nm$^2$. Detailed data are presented in Supplementary S6.

We first investigate $A_G$ in the fabricated devices with various $l_{MIM}$ values. As shown in the inset of Fig. 3a, monolayer graphene extends over the whole MIM region. Figure 3a shows the transmitted intensity of devices with/without graphene as a function of $l_{MIM}$ for $w_{slot}$ = 30 nm and $t_{Au}$ = 20 nm, normalized by that for the Si-wire WG without graphene. By linear fitting, the propagation loss is estimated to be 3.7 dB/μm with graphene and 1.9 dB/μm without graphene. Thus, $A_G$ is 1.7 dB/μm, which is close to the calculated result (2.0 dB/μm). Moreover, we investigated $A_G$ for various $w_{slot}$ values with $t_{Au}$ = 20 nm and 40 nm as summarized in Fig. 3b. The result clearly shows that $A_G$ is greatly enhanced as $w_{slot}$ or $t_{Au}$ decreases. This is exactly as we saw in Fig. 1d, and the measured results are fairly close to the calculated values (solid lines in Fig. 3b). The maximum $A_G$ is 1.7 dB/μm ($w_{slot}$ = 30 nm, $t_{Au}$ = 20 nm), which is 19 times higher than that in previously reported graphene-loaded Si-wire WGs (0.09 dB/μm)[39].

We next proceed to examine the nonlinear response of the graphene absorption. Here, we use an MIM-WG with $w_{slot}$ = 30 nm, $t_{Au}$ = 20 nm, and $l_{MIM}$ = 4 μm, and the entire metal region is covered with monolayer graphene. Picosecond laser pulses (pulse duration ($t_{\text{pulse}}$) of 1.5-2.0 ps, λ = 1550 nm) are injected via the input Si-WG into the sample, and the input pulse energy ($U_{\text{in}}$) dependence of the transmittance is measured (see Methods). Here, $U_{\text{in}}$ is the pulse energy in the input Si-wire WG. As shown in Fig. 4a, the transmittance of the reference Si-wire WG (blue dots) is almost constant up to about 100 fJ, and shows some decrease at larger $U_{\text{in}}$ values owing to the two-photon and free-carrier absorption in Si. However, the transmittance for an MIM-WG with monolayer graphene (black dots) shows a clear increase as $U_{\text{in}}$ increases, which is a typical signature of saturable absorption. This result indicates that the saturable absorption begins at significantly small pulse energy of around 10 fJ.

To analyse the saturable absorption quantitatively, we apply the following attenuation coefficient $\alpha$ to our device,

$$\alpha = \frac{\alpha_S}{1 + P/P_S} + \alpha_{NS} + A_{\text{MIM}}, \quad (1)$$

where $\alpha_S$ and $\alpha_{NS}$ are the saturable and non-saturable absorption coefficients for monolayer graphene, respectively. $P$ and $P_S$ are the optical power intensity and the saturation power, respectively (Supplementary Section S5 for details). $P$ and $P_S$ are replaced with the pulse energy $U$ and saturation energy $U_S$ assuming a constant $t_{\text{pulse}}$. $\alpha_S + \alpha_{NS}$ and $A_{\text{MIM}}$ are deduced from the linear absorption measurement (Fig. 3). $U_S$, $\alpha_S$, and $\alpha_{NS}$ are estimated by fitting the data with Eq. (1). The fitted curve for monolayer graphene is shown in Fig. 4a, and $U_S$ and $\alpha_S$ are estimated to be 12 fJ and 0.63 dB/μm, respectively. The estimated $U_S$ is four orders of magnitude smaller than that in a previously reported graphene-loaded Si-slot WG[40]. For a series of devices with different slot dimensions, we confirmed that the saturable absorption becomes more prominent as the slot dimension decreases (Supplementary Fig. S6), which proved unambiguously that light concentration is the key to reducing the nonlinear operation energy. We also observed essentially the same saturable absorption curves at λ = 1530-1560 nm (Fig. S7). This broadband feature is an important characteristic of our devices, and we attribute it to the nature of the graphene absorbance and the broadband characteristics of our MIM-WG including the mode converter[36].

On the other hand, it is still difficult to obtain an extinction ratio higher than 3 dB with 4-μm-long MIM-WGs. The extinction ratio could be increased by making the MIM-WGs longer, but this also increases the MIM-WG loss. Hence, as another way of increasing the extinction ratio, we load multilayer graphene on the MIM-WG, which does not increase the MIM-WG loss. Based on the parameters obtained by the fitting for the monolayer, we investigate the saturable absorption for multilayer-loaded devices. With the multilayer, $\alpha_S$ and $\alpha_{NS}$ in Eq. (1) are replaced with $N\alpha_S$ and $N\alpha_{NS}$, respectively, where $N$ is the number of graphene layers. The calculated transmittance for bilayer graphene ($N = 2$) is shown as the pink solid line in Fig. 4a, which improves the extinction ratio to over 3 dB.

Motivated by this calculation, we prepared bilayer-graphene-loaded MIM-WGs (see Methods). The red dots in Fig. 4a show the measured transmittance for the bilayer device with $w_{slot}$ = 30 nm, $t_{Au}$ = 20 nm, and $l_{MIM}$ = 4 μm. The result clearly shows saturable absorption with a much larger extinction ratio up to 3 dB, as predicted in the calculation.

Next, we describe all-optical switching experiments with the same bilayer-graphene-loaded MIM-WGs undertaken with the pump-probe method (see Methods). Figure 4b shows the temporal waveform obtained with the pump-probe experiment (Fig. 4c). The blue dots show the autocorrelation of the input pulse ($t_{pulse}$ = 1.8 ps). The pump-probe signal (red dots) shows that the device output is modulated by the input pulse, which demonstrates the expected all-optical switching operation. In this experiment, the control pulse energy in the input Si-wire WG was as small as 71 fJ, which is smaller than that for previous picosecond all-optical switching (see Fig. S1). The observed extinction ratio is 2.2 dB. Interestingly, this ratio is larger than expected from Fig. 4b, and it shows that the whole saturation curve shifts to smaller energies. We observed a similar deviation for all devices (a typical result is shown in Supplementary Fig. S8a), and we discuss its origin in Supplementary S9.

As we pointed out, this type of nonlinear absorption switch is restricted by the recovery time of photo-generated carriers, and this should lead to an asymmetric output waveform. However, the observed output waveform is symmetric in time, and its full width at half maximum (FWHM) is 2.5 ps, which is close to that for the autocorrelation width (2.8 ps) of the input pulse. The reported relaxation time of photo-excited carriers in graphene ranges from 10 fs to several ps[18-21], and our result indicates that the recovery time is much shorter than $t_{pulse}$ (1.8 ps).

Encouraged by this result, we performed the same experiments using a free-space optics setup with a femtosecond laser ($t_{pulse}$ = 210-230 fs, $\lambda$ = 1550 nm; see Methods for measurement details) for the same devices. We first checked the saturable absorption (Fig. 4d). Although the data are scattered, $U_s$ is roughly estimated to be around 3 fJ for the monolayer device (black dots in Fig. 4d), where we assume the same $\alpha_S$ as in the picosecond measurement. A $U_s$ reduction for shorter pulses is naturally expected from the relation $U_s = t_{pulse}P_s$, but this result proves that the recovery time is indeed much shorter than the previous $t_{pulse}$ (1.8 ps). For the bilayer device, the saturation curve becomes slightly steeper, and is also well explained by the calculation.

We next performed a femtosecond pump-probe measurement with $t_{pulse}$ = 210 fs for the bilayer device. As shown in Fig. 5a, we observed clear all-optical switching operation with an extinction ratio of 3.5 dB at a control pulse energy of 35 fJ. Figure 5b shows the extinction ratio for different input pulse energies, demonstrating that an even larger extinction ratio is possible. Note that the switching energy at 2.8 dB is as small as 20 fJ, showing that the switching energy has decreased significantly from that in Fig. 4b (almost scaled to $t_{\text{pulse}}$) when compared at a similar extinction ratio. This switching energy is four orders of magnitude smaller than that for previous graphene-based devices[33], and is the smallest value ever reported for any type of all-optical switching in the picosecond or less regime.

Surprisingly, the output waveform is still symmetric, and the measured FWHM is 370 fs, which is only slightly wider than the autocorrelation width (300 fs) of the input pulse. Assuming the pump-probe signal is given by the correlation between the control and signal pulses, the output pulse width is estimated to be 260 fs, which we regard as the true response time of our all-optical switch. We also predicted the pump-probe signal for $t_{\text{pulse}}$ = 210 fs, and the relaxation time of a graphene carrier $\tau$ is between 100 fs and 1 ps (Fig. 5b). For $\tau >$ 200 fs, the calculated pump-probe signal is clearly broader than the experimental result. Moreover, the calculated signal is asymmetric even at $\tau$ = 100 fs (Supplementary S11). This result indicates that the graphene recovery time is much less than 100 fs, which might be explained by the fast carrier-carrier scattering process[18]. Another possible explanation is ultrafast carrier diffusion out of the 30-nm-wide slot region[41,42]. Since it has been reported that the diffusion time of graphene carriers excited with 1.6-µm laser spot is less than 1 ps[42], much faster diffusion might be possible for our subwavelength devices.

The total insertion loss of the present device is 19 dB in the on-state for an extinction ratio of 3 dB, which consists of the metal absorption, the coupling loss and the graphene absorption. It is noteworthy that this relatively-large insertion loss does not reflect intrinsic limitation of the present device. As discussed in Supplementary S10, the insertion loss is able to be reduced without sacrificing the energy and speed by shortening the device length (by increasing $N$), improving the coupling loss, and improving the saturable component in graphene absorption $\alpha_S/(\alpha_S + \alpha_{NS})$. Assuming the optimized $N$, the ideal coupling loss, and the saturable absorption efficiency reported in other literature[43,44], the insertion loss is estimated to be about 3-5 dB with the same energy and speed shown in Fig. 5.

**Discussion**

In this study, we demonstrate ultrafast all-optical switching with a switching energy of 35 fJ. The extinction ratio is 3.5 dB and the estimated output pulse width is 260 fs. The switching energy is four orders of magnitude smaller than that needed for previous graphene-based devices. A comparison with other types of switches is summarized in Fig. S1, which shows that the switching energy is the smallest of any all-optical switches operating in the picosecond or less regime, and represents the first all-optical switch in the *femto-joule and femto-second* range. In terms of the energy-time product argument, our switch shows a slight improvement over the previous best device, but a detailed analysis of the waveform indicates that the intrinsic response time might be much shorter than 100 fs, suggesting that this device might be able to operate in an unprecedented energy-time product region. In addition to the switching time, the repetition rate is another important feature for optical processing. For example, four-wave mixing devices have demonstrated time-domain mutiplexing at Tbit/s repetition rate for at an energy cost with pico-joule range[45,46]. Although we have not done multiple-pulse experiments at present, we expect that graphene is promising also for ultrafast repetition rate because the reported values of the carrier recombination lifetime are mostly around a few ps or less[24,47-49]. In addition, the fore-mentioned enhancement of the carrier diffusion[42] might accelerate the repetition rate even more in our devices. However, the mechanism of the carrier relaxation in graphene is still far from clear, and more detailed studies, such as the effect of the position of the Fermi level[21], are apparently needed to prove such characteristics. Yet another important feature is that the present device is connected to a conventional Si-WG, and fabricated by conventional processes mostly compatible with Si photonics technologies, which means that this device is easily utilized in Si photonic integrated circuits. The present result has clarified that graphene is superior as an ultrafast and energy-efficient nonlinear material and deep-sub-λ MIM waveguides is a superior device platform, if both are appropriately combined, thus showing the way to fully utilize the advantages of hybrid nanomaterial-nanophotonic platforms for energy-efficient high-speed processing. We expect that our energy-efficient high-speed optical switches will be implemented in a processor chip integrated with photonic integrated circuits, such as a photonic network-on-chip, where chip-scale communications or routing will be needed. Furthermore, an energy-efficient and ultrafast optical nonlinearity unit is required in a nanophotonic neural network chip, and the optical nonlinearity of graphene has attracted special attention[50,51]. We believe that the importance of our device will be further increased with the progress of architectures for on-chip photonic processing.

FIGURES

Figure 1| Schematic and simulation of a graphene-loaded MIM-WG for all-optical control. **a**, Schematic of the graphene-loaded MIM-WG. **b**, Cross-sectional side view of the MIM-WG at the broken red line shown in **a**. **c**, Calculated field profile ($|E|^2$) of the eigenmode of the graphene-loaded MIM-WG. The air slot width $w_{slot}$ is 30 nm, the Au thickness $t_{Au}$ is 20 nm, and λ = 1550 nm. Scale bar, 20 nm. **d**, Calculated absorption coefficient of graphene $A_G$ for various $w_{slot}$ and $t_{Au}$ values. The solid grey line shows $A_G$ value for the graphene-loaded Si-wire WG (the core size is 400 nm × 200 nm). **e**, Field enhancement factor *EF* for various $w_{slot}$ and $t_{Au}$ values. The field intensity ($|E|^2$) was averaged over the graphene position in the MIM-WG slot, and *EF* was defined as the ratio of the average field intensity against that with the Si-wire WG structure.

Figure 2| Scanning electron microscope (SEM) images of the fabricated samples. **a**, SEM image around the mode converter for an air slot width $w_{slot}$ of 70 nm and an Au thickness $t_{Au}$ of 40 nm. Scale bar, 1 μm. **b**, Top view of schematic of the graphene-loaded MIM-WG. **c**,**d**, SEM images of the graphene-loaded MIM-WGs with $w_{slot}$ = 70 nm, $t_{Au}$ = 40 nm, and graphene lengths $l_G$ of 3.2 μm (**c**) and 7.2 μm (**d**). The MIM-WG length $l_{MIM}$ was 10 μm. Scale bar, 2 μm.

Figure 3| Measured linear absorption of graphene-loaded MIM-WGs. **a**, The MIM-WG length $l_{MIM}$ dependence of the average relative transmittance in the MIM-WGs with and without graphene. The air slot width $w_{slot}$ and the Au thickness $t_{Au}$ were 30 nm and 20 nm, respectively. **b**, Measured and calculated absorption coefficient of graphene $A_G$ for various $w_{slot}$ and $t_{Au}$ values. We obtained $A_G$ from the measured propagation loss with

and without graphene for $t_{Au}$ = 20 nm. For $t_{Au}$ = 40 nm, we changed the graphene length with a fixed $l_{MIM}$ to estimate $A_G$.

Figure 4| Saturable absorption and demonstration of all-optical switching. **a**, Saturable absorption with picosecond laser pulses in the graphene-loaded (monolayer, bilayer) MIM-WGs and the reference Si-wire WG (without graphene). The air slot width $w_{slot}$, the Au thickness $t_{Au}$, and the MIM-WG length $l_{MIM}$ were 30 nm, 20 nm, and 4 μm, respectively. λ of the input pulse was 1550 nm. **b**, All-optical switching with the pump-probe method using picosecond laser pulses in a bilayer-graphene-loaded MIM-WG with $w_{slot}$ = 30 nm, $t_{Au}$ = 20 nm, and $l_{MIM}$ = 4 μm (red dots). The autocorrelation of the input pulse is also plotted (blue dots). The input pulse is explained with a sech$^2$ function and its pulse width was 1.8 ps. The pump-probe signal was fitted with the autocorrelation of a sech$^2$ function (grey solid line). The full width at half maximum of the pump-probe signal (2.5 ps) is slightly wider than the autocorrelation width (2.8 ps) of the input pulse, and it is expected that the input pulse was narrowed by the self-phase modulation in the optical fibre. The control and signal pulse energies were 71 fJ and 10 fJ, respectively, in the input Si-wire WG. **c**, Schematic illustration of the pump-probe measurement. **d**, Saturable absorption with femtosecond laser pulses in the same samples as in **a**. λ of the input pulse was 1550 nm.

Figure 5| Demonstration of all-optical switching with femtosecond laser pulses. **a**, All-optical switching in the bilayer-graphene-loaded MIM-WG with $w_{slot}$ = 30 nm, $t_{Au}$ = 20 nm, and $l_{MIM}$ = 4 μm (red dots). The autocorrelation of the input pulse is also plotted (blue dots). The input pulse is explained with a Gaussian function, and its pulse width $t_{pulse}$ was 210 fs. The pump-probe signal was fitted with a Gaussian function (black solid

line). The control and signal pulse energies were 35 fJ and 1.3 fJ, respectively, in the input Si-wire WG. The pump-probe signals predicted from $t_{pulse}$ (210 fs) and the relaxation time of the graphene carrier $\tau$ are also shown, where two Gaussian functions (FWHM = 210 fs) and single exponential decay function ($\tau$ is from 100 fs to 1 ps) are convoluted. The pink, orange yellow, green, light blue solid lines are for $\tau$ = 100 fs, 200 fs, 300 fs, 500 fs, and 1 ps. **b**, Control pulse energy dependence of extinction ratio for bilayer-graphene-loaded MIM-WG ($w_{slot}$ = 30 nm, $t_{Au}$ = 20 nm, $l_{MIM}$ = 4 μm).

**METHODS**

**Numerical calculation**

We calculated the eigenmodes of the graphene-loaded MIM-WGs with the finite element method (COMSOL Multiphysics) in order to estimate $A_G$ in the graphene-loaded MIM-WGs. The eigenmodes of the WGs each have a complex effective index, and its imaginary part gives the propagation length $L$ (Supplementary Section S3). Here, we calculated $1/L$ with and without graphene, and $A_G$ was obtained from the difference between these two values. In the modelling, we introduced graphene as the boundary condition. The surface current density induced in graphene $\mathbf{J}_s$ is given by its conductivity $\sigma$ and the tangential component of the electric field $\mathbf{E}_\tau$ as $\mathbf{J}_s = \sigma \mathbf{E}_\tau$. $\sigma$ is given by the Kubo formula[52], and we assumed a chemical potential of 0.2 eV, a temperature of 300 K, and a relaxation time of 100 fs (Supplementary Section S2). We took the permittivities of Au ($-116+12.1i$), Si (12.1), and $SiO_2$ (2.09) from previous studies[53,54].

**Sample fabrication**

First, we fabricated Si-wire WGs on silicon-on-insulator (SOI) substrates. We prepared SOI wafers with top Si layer thicknesses of 200 and 220 nm. The wafer with a top Si layer thickness of 200 nm was used for a 20-nm-thick Au layer to obtain better coupling efficiency between a Si-wire WG and an MIM-WG. The other wafer was used for a 40-nm-thick Au layer. The resist patterns for the Si-wire WGs were formed by electron beam (EB) lithography (an acceleration voltage of 125 kV), and the WG patterns were transferred by inductively coupled plasma etching. The fabricated Si-wire WG was 400 nm wide, and it was connected to a 3-μm-wide Si-WG, which was used to input (output) light from (to) an optical fibre or free space. The taper length for the plasmonic mode converter was 600 nm. Next, we fabricated MIM-WGs on the substrate. The resist patterns were formed by EB lithography, and a 20-nm or 40-nm Au layer was evaporated after a 1-nm Ti layer with an EB evaporator. Then, the metal patterns were fabricated with a lift-off technique. We fabricated MIM-WGs with $w_{\text{slot}}$ = 30-50 nm for $t_{\text{Au}}$ = 20 nm and $w_{\text{slot}}$ = 60-100 nm for $t_{\text{Au}}$ = 40 nm. Next, we loaded graphene on the MIM-WGs. We prepared monolayer graphene grown on copper foil by chemical vapour deposition. The polymethyl methacrylate (PMMA) layer (200 nm) was spin-coated on the graphene, and the copper foil was etched in ammonium persulfate solution. The rinsed PMMA/graphene film was floated on deionized water and was transferred to the SOI wafers. After drying the SOI wafers, we baked them at 180°C. Then, we removed the PMMA. When we fabricated the bilayer-graphene-loaded MIM-WGs, we transferred the monolayer graphene twice. Then, we patterned the loaded graphene by reactive ion etching with $O_2$ plasma, using a 1-μm-thick PMMA pattern. Finally, we cleaved the wafer on the wide Si-WGs.

**Linear absorption measurement**

$A_G$ was investigated by measuring the transmittance of the fabricated graphene-loaded MIM-WGs. We used a continuous-wave λ-tuneable laser, and the laser light was input into the 3-μm-wide Si-WGs in the samples via the optical fibre. The transmitted light was output from the 3-μm-wide Si-WGs and connected to the optical fibre. Then, the output power was detected. The laser light was input in the TE mode (λ = 1500-1600 nm), and the input power was 10-100 μW. We used a low input power to avoid nonlinear optical effects such as the saturable absorption in graphene and the two-photon absorption in Si-WGs. With the reference Si-wire WG, where the only 400-nm-wide Si-wire WG was connected to the 3-μm-wide Si-WGs, the insertion loss between the optical fibres was typically about 17 dB, and the power was about 1-10 μW in the middle of the samples. Then, we obtained the average relative transmittance in the measured wavelength range (the reference was the transmittance of the Si-wire waveguide without graphene).

**Saturable absorption measurement**

To induce saturable absorption, we input picosecond or femtosecond laser pulses. For the picosecond laser pulses, we used a mode-locked fibre laser ($t_{pulse}$ = 1.5-2.0 ps, $\lambda$ = 1550 nm). The original repetition rate was 10 MHz, which was changed to 0.1-10 MHz with a pulse picker. For a larger energy, the repetition rate was reduced to reduce the average power. The laser pulses were input into the samples via the optical fibre, and the output light was also connected to the optical fibre. Then, the output power was detected (Supplementary Fig. S12a). For femtosecond laser pulses, we used an optical parametric oscillator (OPO) with a mode-locked Ti:sapphire laser. The repetition rate was 80 MHz, and $t_{pulse}$ = 210-230 fs at $\lambda$ = 1550 nm. The laser pulses were directly input into the input Si-WGs in the samples from free space with an objective lens. The output light was connected to the optical fibre. $U_{in}$ was defined as the pulse energy in the input Si-wire WG, which was estimated by measuring the transmittance of the reference Si-wire WG. $U_{in}$ is the pulse energy in the middle of the samples, and the MIM WGs were placed about 200 μm behind the middle of the samples.

**Picosecond pump-probe measurement**

In the picosecond pump-probe measurement, we used the mode-locked fibre laser that we used in the saturable absorption measurement, and set the repetition rate at 1 MHz. The laser light was split into control (pump) and signal (probe) lights, and the time delay between the control and signal pulses was changed. The signal light was modulated at a frequency of 1.1 kHz with an acousto-optic modulator. Then, the control and signal lights were combined and input into the samples via the optical fibre. The output light was connected to the optical fibre, and detected with a photo-receiver. Then, we measured the output power of the signal light with a lock-in amplifier for different time delays (Supplementary Fig. S12b).

**Femtosecond pump-probe measurement**

In the femtosecond pump-probe measurement, we used the OPO that we used in the saturable absorption measurement. The laser light was split into control and signal lights with a beam splitter in free space, and the time delay between the control and signal pulses was changed. The signal light was modulated at a frequency of 137 Hz with a mechanical chopper for the lock-in detection. In addition, we inserted a piezo mirror in the optical path of the signal light to suppress the interference between the control and signal lights. We applied a voltage of 100 V (displacement is about 10 μm) at a frequency of 2 kHz. Then, the control and signal lights were combined and input into the samples via the objective lens. We then measured the output power in the same way as in the picosecond pump-probe measurement (Supplementary Fig. S12c).

ACKNOWLEDGEMENTS

We thank E. Kuramochi and T. Tamamura for support with the fabrication, K. Takata for support with the measurement, and A. Shinya, N. Matsuda, and Y. Ogawa for fruitful discussions. This work was supported by JSPS KAKENHI Grant Number JP15H05735.

AUTHOR CONTRIBUTIONS

M.O. designed and fabricated the sample, performed the experiment, analysed the data, and wrote the manuscript. M.H. and M.T. numerically designed and fabricated the sample, performed the experiment, and analysed the data. K.N. supported the measurement setup and the discussion. H.S. and H.C. supported the graphene process and the discussion. M.N. conceived the work, designed the sample, analysed the data, wrote the manuscript, and led the project.

Figure 1

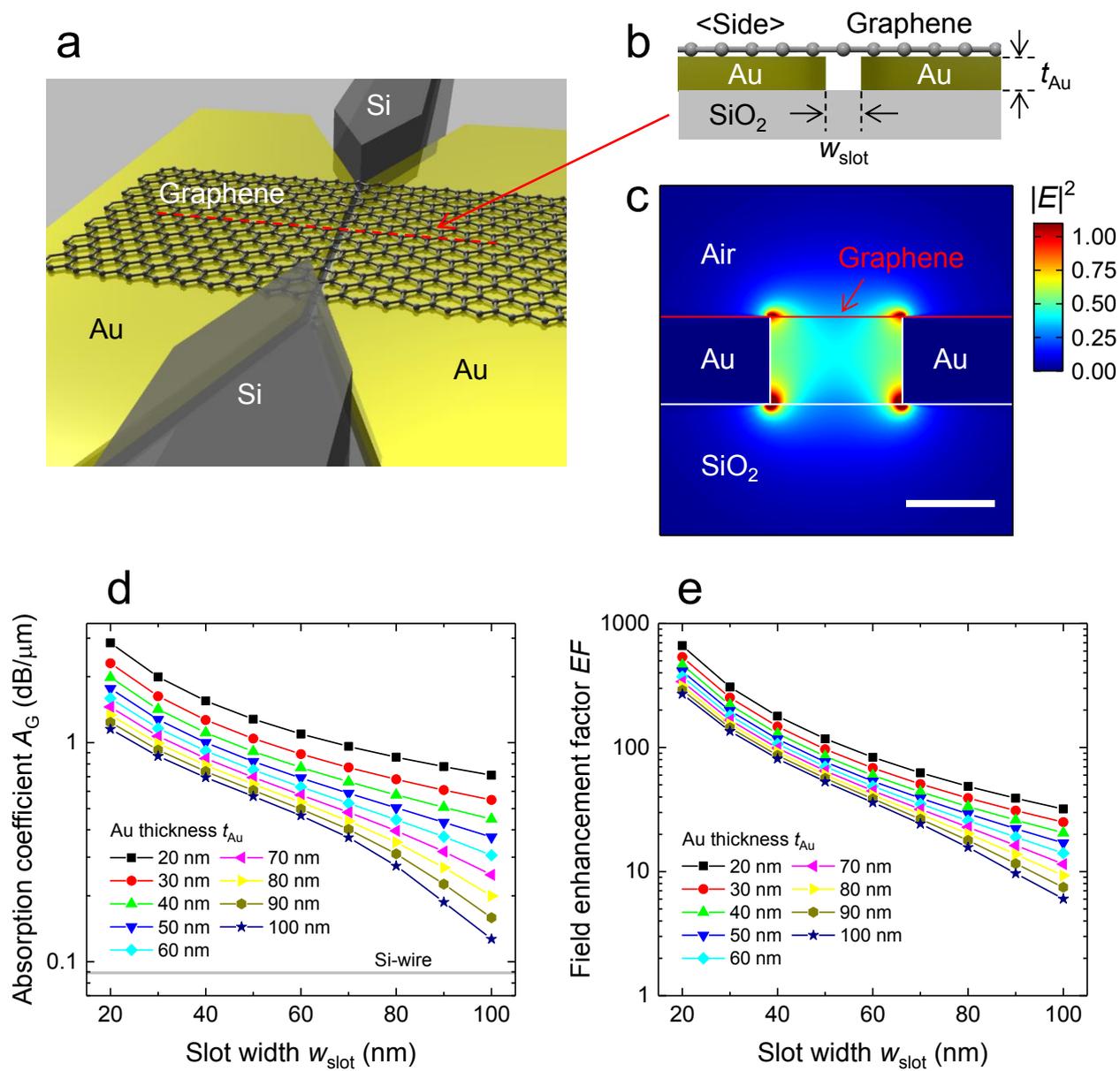

Figure 2

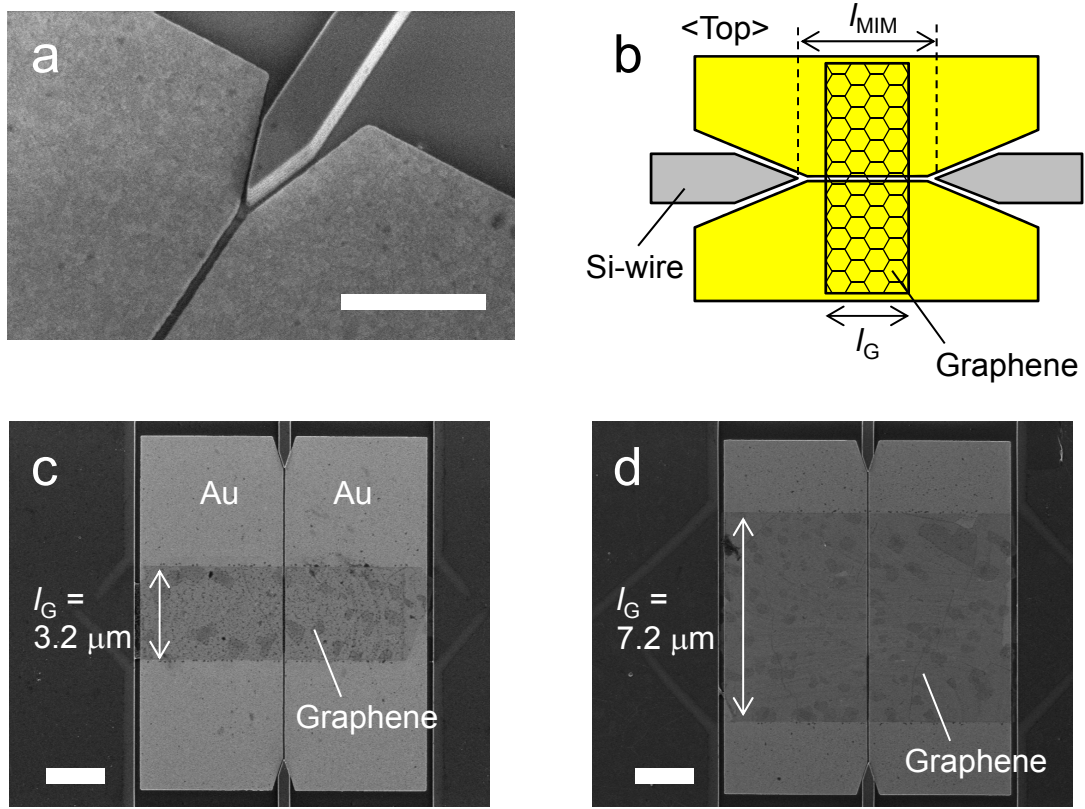

Figure 3

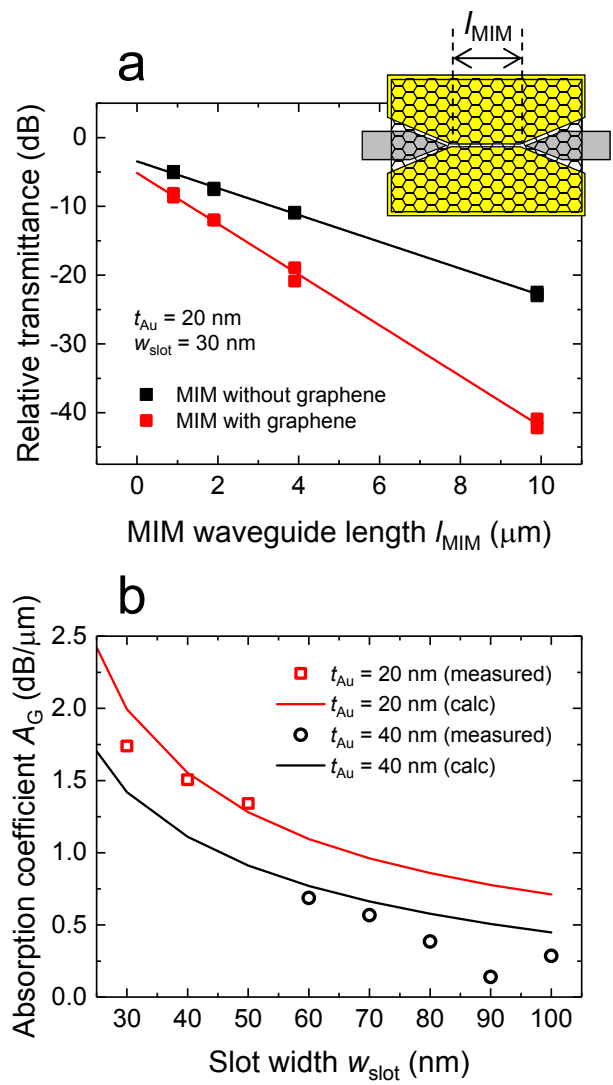

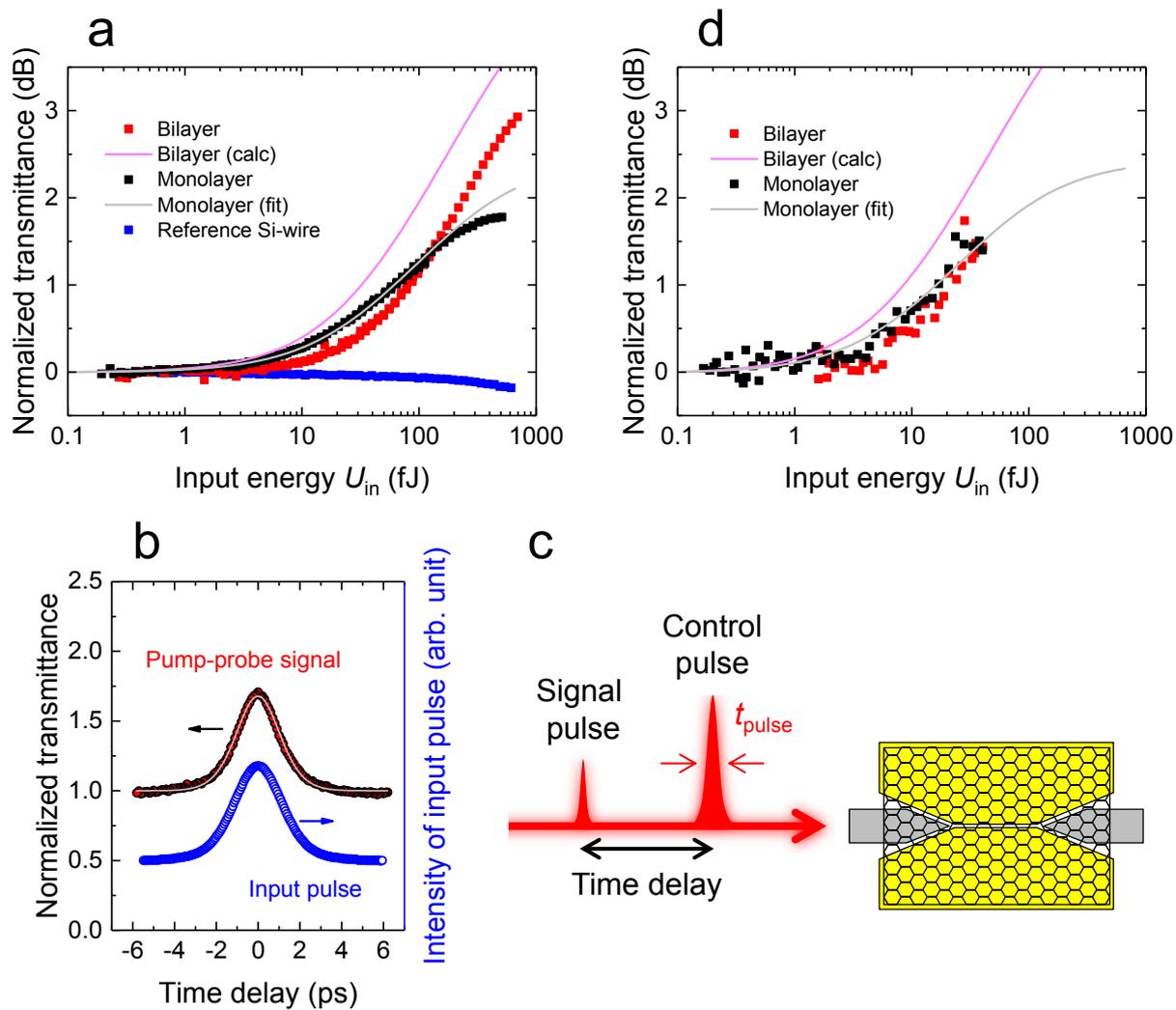

Figure 5

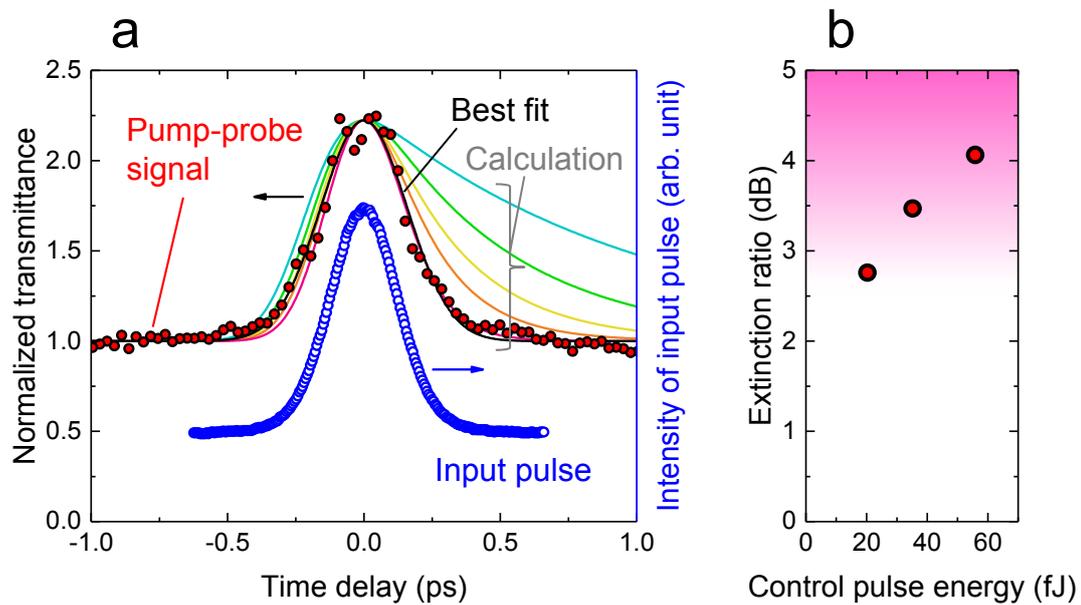